\documentclass[fleqn, usenatbib]{mn2e}
\usepackage{amssymb,amsmath}
\usepackage{graphicx,epstopdf}
\usepackage{xcolor,epstopdf}
\usepackage{multirow}
\usepackage[T1]{fontenc}
\usepackage{ae,aecompl}

\title[Variability of double-peaked emitters]
{Properties of long-term optical variability of active galactic nuclei with 
     double-peaked broad low-ionization emission lines}
\author[Zhang \& Feng]
       {Xue-Guang Zhang\thanks{Corresponding author Email:
            zhangxg23@sysu.edu.cn}, \& Long-Long Feng\\
       Institute of Astronomy and Space Science, Sun Yat-Sen University, 
            Guangzhou, 510275, P. R. China}

\date{}

\begin{document}
\pagerange{\pageref{firstpage}--\pageref{lastpage}} \pubyear{2016}
\maketitle
\label{firstpage}

\begin{abstract}
    In this manuscript, we study properties of long-term optical 
variability of a large sample of 106 SDSS spectroscopically confirmed 
AGN with double-peaked broad low-ionization emission lines (double-peaked 
emitters). The long-term optical light curves over 8 years are collected 
from the Catalina Sky Surveys Data Release 2. And, the Damped Random Walk 
(DRW) process is applied to describe the long-term variability of the 
double-peaked emitters. Meanwhile, the same DRW process is applied to 
long-term optical light curves of more than 7000 spectroscopically confirmed 
normal quasars in the SDSS Stripe82 Database. Then, we can find that the DRW 
process determined rest-frame intrinsic variability timescales 
$\ln(\tau/{\rm days})$ are about 5.8 and about 4.8 for 
the double-peaked emitters and for the normal quasars, 
respectively. The statistically longer intrinsic variability timescales 
can be confirmed in the double-peaked emitters, after 
considerations of necessary effects, such as the effects from different 
distributions of redshift, BH mass and accretion rate between the 
double-peaked emitters and the normal quasars. Moreover, a radial 
dependence of accretion rate $\dot{m}_{\rm R}~\propto~R^\beta$ with 
larger values of $\beta$ could 
be an acceptable interpretation of the longer intrinsic variability 
timescales in the double-peaked emitters. Therefore, there are 
different intrinsic properties of emission regions between the double-peaked 
emitters and the normal 
quasars. The double-peaked emitters can be well treated 
as an unique subclass of AGN. 
\end{abstract}


\section{Introduction}
 
   Variability is one of fundamental characteristics of active galactic 
nuclei (AGN) \citep{um97}. To study AGN variability can 
provide further clues to properties of emission regions of 
AGN \citep*{re84, tc00, haw02, hh06, me11}. And commonly, there are 
three characteristic timescales of around days to hundreds of days 
through accretion physics \citep*{pe01, kbs09}, the light crossing 
timescale ($t_{\rm lc}$), the gas orbital timescale ($t_{\rm orb}$) and 
the accretion disk thermal timescale ($t_{\rm th}$), 
\begin{equation}
\begin{split}
&t_{\rm lc}~=~1.1~\times~M_8\times~R_{\rm 2}~{\rm days} \\
&t_{\rm orb}~=~104~\times~M_8\times~R_{\rm 2}^{3/2}~{\rm days}\\
&t_{\rm th}~=~4.6~\times~(0.01/\alpha)~\times~M_8\times~R_{\rm 2}^{3/2}~{\rm years}
\end{split}
\end{equation}, 
where $M_8~=~M_{\rm BH}/{\rm 10^8~M_\odot}$ represents black hole (BH) 
mass, $R_{\rm 2}$ represents distance of emission regions to central 
engine in unit of ${\rm 100~R_G}$ (${\rm R_G}$ is the Schwarzschild 
radius), and $\alpha$ means the standard disk viscosity parameter.  
Therefore, the expected AGN variability timescales 
strongly connected with accretion physics should sensitively depend 
on the fundamental parameter of AGN, the BH mass, and on the distance 
of emission regions to the central black hole. Hence, to study AGN 
variability could provide further properties of emission 
regions of AGN.

    More recently, the Damped Random Walk (DRW) process \citep{bd02} 
(or AutoRegressive process), with two basic parameters of the intrinsic 
variability timescale $\tau$ and the intrinsic variability amplitude 
$\sigma$, has been proved to be a preferred modeling 
process to describe AGN intrinsic variability. \citet{kbs09} firstly 
proposed the DRW process to describe the AGN intrinsic variability, and 
found that the AGN intrinsic variability timescales are consistent with 
disk orbital or thermal timescales. \citet{koz10} provided an improved 
robust mathematic method to estimate the DRW process parameters, and 
found that AGN variability could be well modeled by the DRW process. 
Then, \citet{zu11} provided a public code of JAVELIN 
(http://www.astronomy.ohio-state.edu/\~{}yingzu/codes.html\#javelin)  
(Just Another Vehicle for Estimating Lags In Nuclei) based on the 
method in \citet{koz10} to describe the AGN variability by the DRW 
process. Meanwhile, there are many other reported studies on the AGN 
variability through the DRW process. \citet{mi10} modeled the 
variability of about 9000 spectroscopically confirmed quasars covered 
in the SDSS (Sloan Digital Sky Survey) Stripe82 region, and found 
correlations between the AGN parameters and the DRW process determined 
parameters. \citet{bj12} proposed an another fully probabilistic 
method for modeling AGN variability by the DRW process. \citet{ak13} 
have shown that the DRW process is preferred to model AGN variability, 
rather than several other stochastic and deterministic models, by fitted 
results of long-term variability of 6304 quasars. \citet{zu13} have 
checked that the DRW process provided an adequate description of 
AGN optical variability across all timescales. Therefore, the DRW 
process determined parameters from the long-term AGN variability can 
be well used to check or predict further different properties of 
different kinds of AGN with probable different intrinsic properties 
of emission regions.

   Among broad line AGN, there is one special kind of AGN, the AGN 
with double-peaked broad low-ionization emission lines (hereafter, 
double-peaked emitters). Since the first reported 
double-peaked emitter of 3C390.3\ in 1980s \citep*{ss83}, 
more and more double-peaked emitters have been reported 
in \citet*{eh94, st03, sh11}, etc.. And the proposed theoretical model 
with broad emission lines coming from central accretion disk has been 
preferred to explain the unique double-peaked broad emission lines 
\citep*{ch89, el95, fe08}. Long-term variability of a few 
double-peaked emitters, especially profile variability of the 
double-peaked broad emission lines, have been reported in \citet*{era97, 
sh01, sn03, lew10, zh13}, etc., and applied to test the accretion 
disk origin of the double-peaked broad emission lines. 

   However, there are so far no clear statistic results on the 
long-term variability properties of double-peaked emitters  
by the DRW process. As the results in \citet*{mi10, sch10, mh11, sch12, 
zh14}, etc., the public SDSS Stripe82 Database (hereafter, SDSS S82) 
\citep{bra08} is the widely used database to study the long-term SDSS 
AGN variability. However, there are only several 
double-peaked emitters in the SDSS S82, and it is hard to do 
statistical research on the long-term variability of a large sample 
of double-peaked emitters through the SDSS S82. 
More fortunately, the well-known Catalina Sky Survey (CSS) \citep{dra09} 
has provided public long-term (over 8 years) light curves of more 
than 500 million objects from an area of 33000 square degrees. And we 
can find that hundreds of double-peaked emitters are 
covered in the CSS. Thus, it is the time to statistically study the 
long-term variability of a large sample of double-peaked 
emitters, to check whether are there different properties of 
emission regions in the double-peaked 
emitters.

  The manuscript is organized as follows. In Section 2, we presented 
our main results and necessary discussions on long-term variability 
properties determined by the DRW process for a large sample of 
double-peaked emitters and a large sample of normal 
quasars. Then, in Section 3, we gave our final conclusions.

\section{Main Results and Discussions}

\begin{figure*}
\centering\includegraphics[width = 18cm,height=16cm]{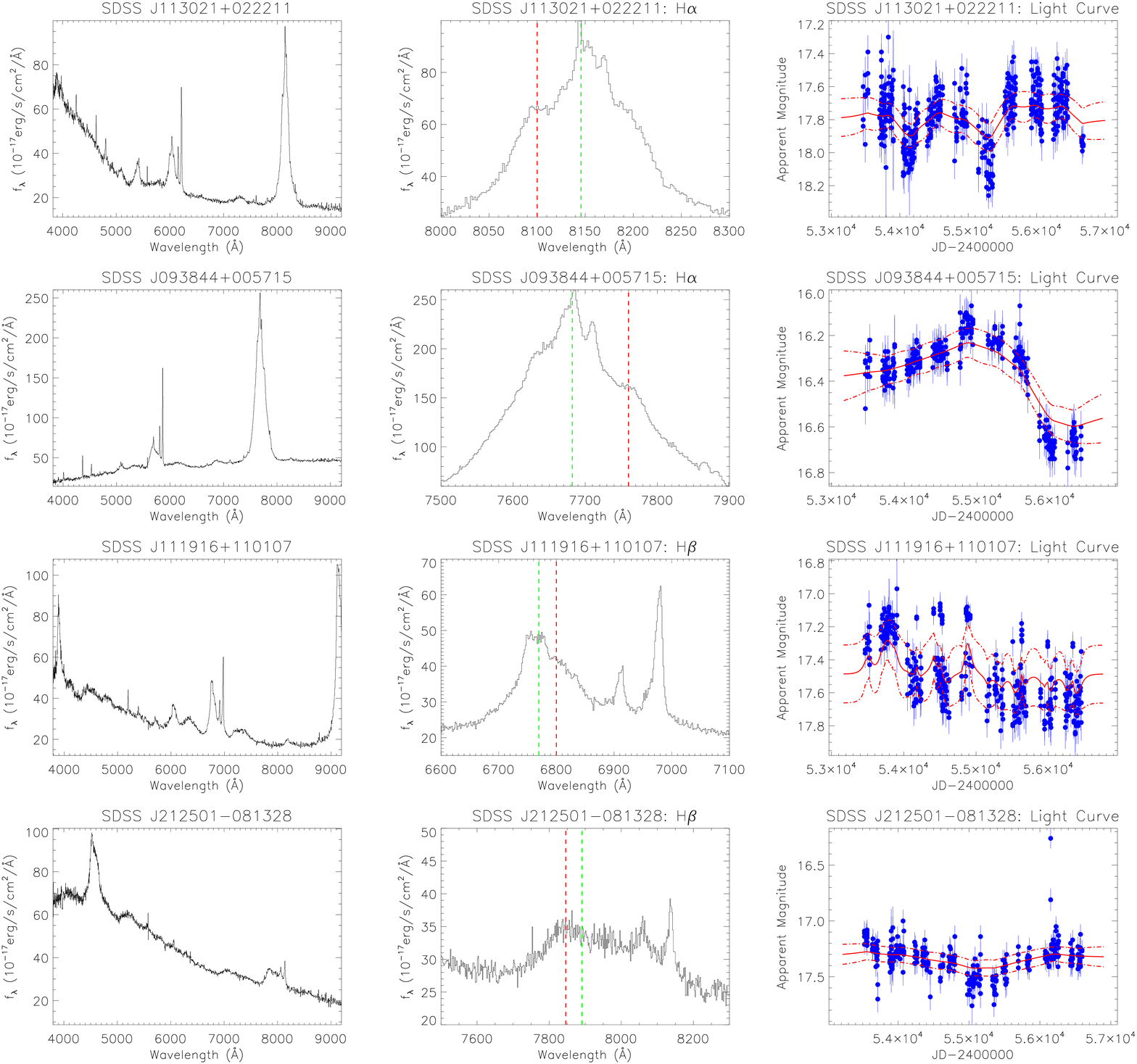}
\caption{SDSS spectra and long-term light curves of four 
double-peaked emitters as examples. Left panels show the SDSS spectra.
Middle panels show the double-peaked broad H$\alpha$ or double-peaked
broad H$\beta$ with marked positions of clear shoulders in red dashed
lines and of center wavelengths of narrow H$\alpha$ or narrow H$\beta$
in green dashed lines. And right panels show the light curves from the
CSS DR2. In the right panels, solid circles in blue with 
error bars are for the observed light curves, solid lines 
and dot-dashed lines in red show the best-fitted results by the DRW 
process and the corresponding 1sigma confidence bands, respectively.
}
\label{dbp}
\end{figure*}

    The first large sample of more than 100 double-peaked 
emitters in SDSS was reported in \citet{st03}. And more recently, the 
systematically determined large sample of more than 200 
double-peaked emitters in SDSS quasars was reported in \citet{sh11}. 
In this manuscript, the spectroscopically confirmed 
double-peaked emitters in \citet{sh11} and in \citet{st03} are 
collected and created as our main parent sample. Here, 
the double-peaked emitters are selected from the catalogue of 
\citet{sh11} by the following criterion on the key parameter of 
$special\_interest\_flag$: $special\_interest\_flag~\land~2L^0~NE~0$ (see 
http://quasar.astro.illinois.edu/BH\_mass/dr7.htm\#special\_flag).The 
double-peaked emitters were classified as such by \citet{sh11} based 
on the profile of the Balmer lines. And, in \citet{sh11}, the selected 
objects based on the criterion above are called disk emitters with their 
double-peaked broad emission lines probably expected from central 
accretion disks. However, we should note that binary black hole model 
\citep{ga96, zh07} or bipolar outflow model \citep{zh91} could also 
lead to double-peaked broad emission lines. Hence, the term 'double-peaked 
emitters' rather than the term 'disk emitters' is used in the manuscript 
to represent the AGN with double-peaked broad emission lines. Here, 
we do not show the spectra of all the objects in the parent sample, 
but SDSS spectra of four double-peaked emitters as examples are shown 
in the left and the middle panels of Fig.~\ref{dbp}.

    Then, light curves of all the spectroscopically 
confirmed double-peaked emitters in SDSS (220 objects from \citet{sh11} 
and 116 objects from \citet{st03}) are searched in the Catalina Sky 
Surveys Data Release 2 (CSS DR2), with position radius within 1\arcsec. 
There are 256 double-peaked emitters with available light curves 
provided by the CSS DR2. More detailed descriptions on the instruments 
and the techniques of CSS can be found in \citet{dra09} and in 
\citet{lar03}, and in the webpage of 
http://nesssi.cacr.caltech.edu/DataRelease/. Before listing the 
double-peaked emitters, further properties of long-term 
variability should be firstly checked by the DRW process, because light 
curves of part of objects can not be well described by the DRW process 
due to large uncertainties (smaller signal-to-noise ratio) of the CSS 
DR2 provided light curves (see an example shown in the bottom right 
panel of following Fig.~\ref{s82_css}).

   The public code of JAVELIN provided by \citet{zu11} and \citet{ zu13} 
is applied to describe the CSS DR2 provided long-term variability of 
the double-peaked emitters based on the DRW process. 
When the JAVELIN code is applied, through the MCMC (Markov Chain Monte 
Carlo) analysis with the uniform logarithmic priors of the DRW process 
parameters of $\tau$ and $\sigma$ covering every possible corner of the 
parameter space ($0~{\rm days}<~\tau~<~1e+5~{\rm days}$ and 
$0~{\rm mag/day^{0.5}}<~\sigma~<~1e+2~{\rm mag/day^{0.5}}$), the posterior 
distributions of the DRW process parameters can be well determined and 
provide the final accepted parameters and the corresponding statistical 
confidence limits. Here, when the JAVELIN is applied, the light curves 
from the CSS DR2 are firstly re-sampled with one mean 
value for multiple observations per day. Then, based on the DRW process 
determined parameters of $\ln(\tau)$ and $\ln(\sigma)$ at least 5 times 
larger than the corresponding uncertainties, 106 spectroscopically 
confirmed double-peaked emitters are finally collected 
and created as our main sample.

    Here, we do not show the SDSS DR2 provided light curves of all 
the 106 double-peaked emitters, but right panels of 
Fig.~\ref{dbp} shows four examples on the light curves and the 
corresponding best fitted results by the DRW process. Table~1 lists the 
basic parameters of the 106 double-peaked 
emitters with redshift less than 0.62. Here, the listed values of 
$\ln(\tau)$ and $\ln(\sigma)$ are the rest-frame values after 
considerations of the cosmological time dilation as discussed in 
\citet{kbs09}. The mean values of $\ln(\tau/{\rm days})$ 
are about 5.77 and 6.02 for the double-peaked emitters, 
in the rest-frame and in the observed-frame, respectively. Before 
proceeding further, we can find that \citet{mi10} have reported the 
mean value of the observed-frame $\ln(\tau/{\rm days})~\sim~5.29$ 
($\log(\tau/{\rm days})~\sim~2.3$) for about 9000 quasars in the SDSS 
S82. Therefore, more longer $\ln(\tau)$ could be roughly found in the 
double-peaked emitters than in the normal quasars.

   Then, besides the DRW process determined properties of the long-term
variability of the 106 double-peaked emitters, it is
necessary to determine long-term variability properties for a large 
sample of normal broad line AGN, in order to find more clearer variability
difference between the double-peaked emitters and the
normal broad line AGN (or normal quasars). Here, all the spectroscopically
confirmed 21888 quasars in the SDSS S82 are considered (see also our 
previous results on quasars in SDSS S82 in \citet{zh14}). The long-term
light curves of the quasars are collected from the SDSS S82. The same 
procedure is applied to fit the SDSS g-band light curves of the quasars. 
Here, the main reason to consider the SDSS g-band light curves of 
the normal quasars is mainly due to the following reason. The CSS DR2 
provided magnitudes are similar as the standard V band magnitudes, 
meanwhile, the SDSS g-band magnitudes are sensitively proportional to 
standard V band magnitudes \citep{je05}. Then, 7804 normal quasars 
are collected with reliable values of $\ln(\tau)$ and $\ln(\sigma)$ 
(at least 5 times larger than the corresponding uncertainties).

\begin{figure*}
\centering\includegraphics[width = 16cm,height=10cm]{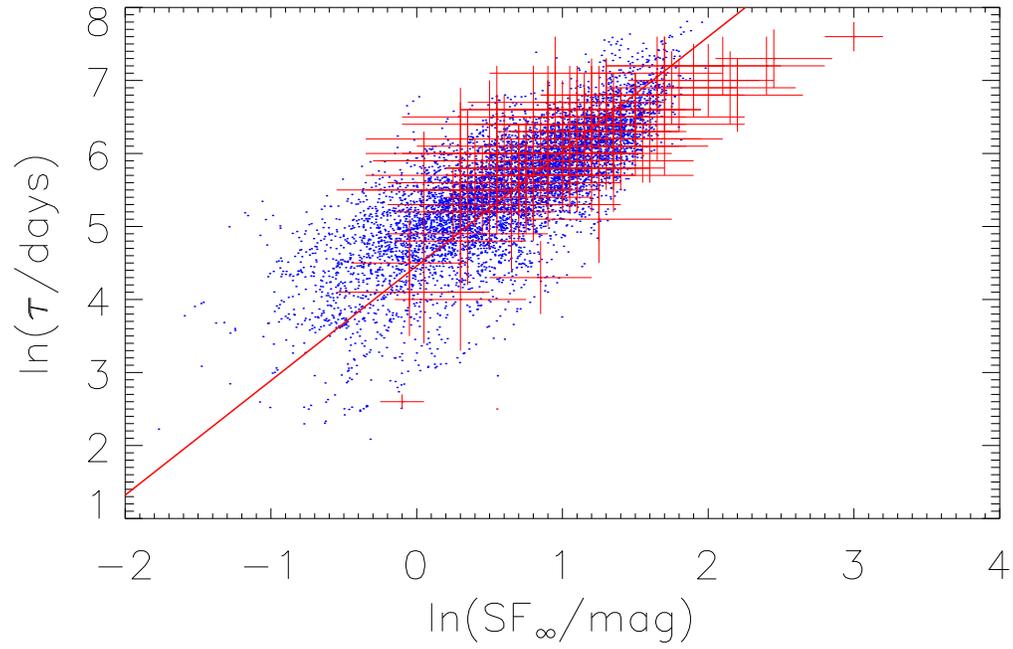}
\caption{
Correlation between the observed-frame $\ln(\tau)$ and the observed-frame 
$\ln(SF_{\infty})$ for the normal quasars (in blue dots) and for the 
double-peaked emitters (in red dots with error bars). The solid line
in red shows the best fitted result of $\tau\propto SF_{\infty}^{\sim1.57}$.
}
\label{drw_ori}
\end{figure*}

\begin{figure*}
\centering\includegraphics[width = 16cm,height=20cm]{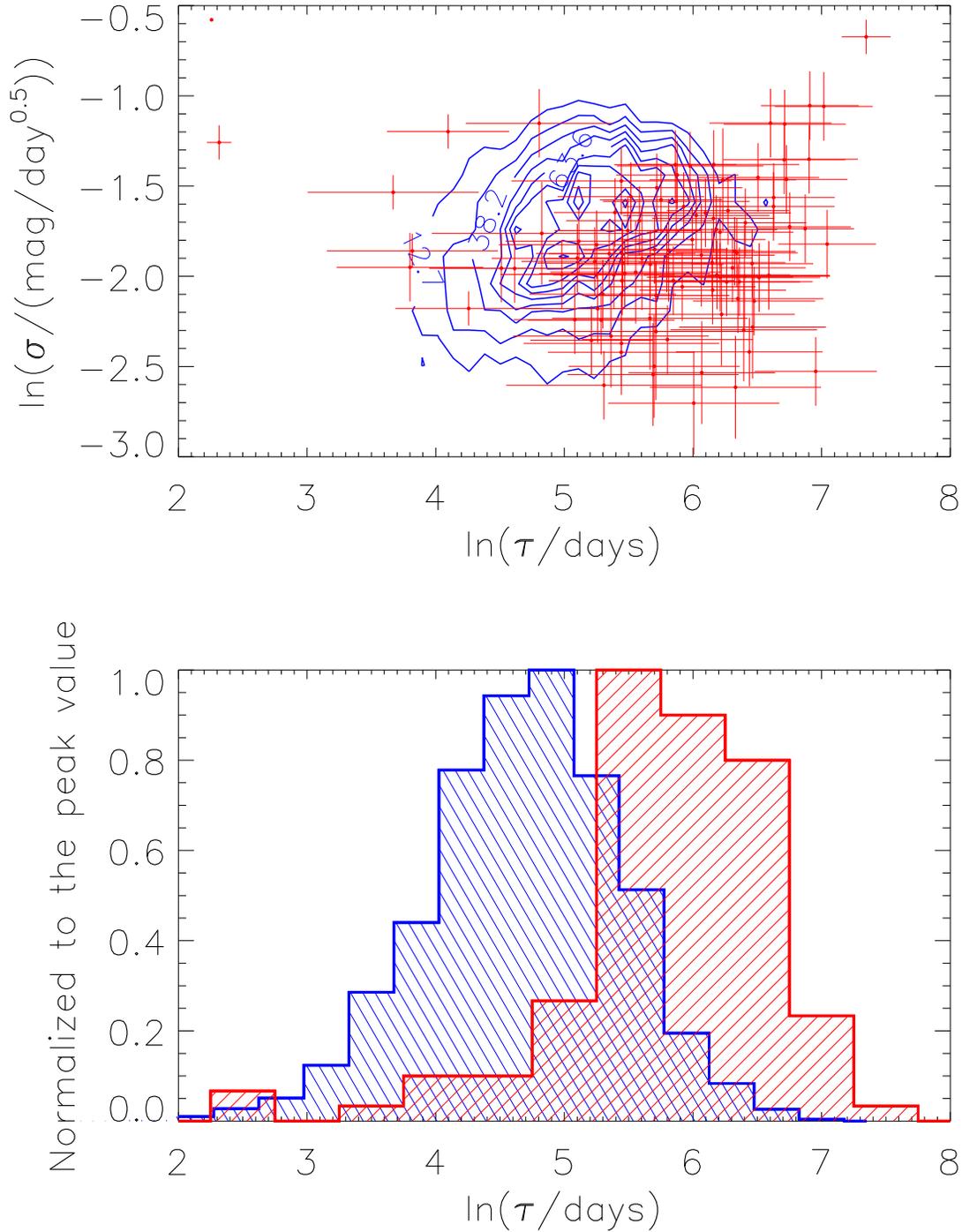}
\caption{Properties of the DRW process determined parameters
of the rest-frame $\ln(\tau)$ and the rest-frame $\ln(\sigma)$ of the
double-peaked emitters and the normal quasars.
Top panel shows correlation between the rest-frame $\ln(\tau)$ and the
rest-frame $\ln(\sigma)$ for the double-peaked emitters 
in solid circles in red, for all the normal quasars in contour with blue 
lines, respectively. Bottom panel shows distributions of the rest-frame
$\ln(\tau)$ for the double-peaked emitters in red and for
the normal quasars in blue, respectively.}
\label{drw}
\end{figure*}

   Before proceeding further, in order to ensure 
confidence in our measured parameters of $\ln(\tau)$ and $\ln(\sigma)$, 
Fig.~\ref{drw_ori} shows the correlation between the observed-frame 
$\ln(\tau)$ and the observed-frame 
$\ln(SF_{\infty})~=~\ln(\sigma\times\sqrt{\tau})$\footnote{In \citet{mi10}, 
the parameter $SF_{\infty}~=~\sigma\times\sqrt{\tau}$ was applied.} 
which could be described by a power law function with slope of 
$\sim1.3$ as discussed in \citet{mi10} for the quasars in the SDSS S82 
(see Figure 3\ in \citet{mi10}). It is clear that through our measured 
parameters of $\ln(\tau)$ and $\ln(\sigma)$ for the normal quasars, 
there is a strong linear correlation between the observed-frame 
$\ln(\tau)$ and the observed-frame $SF_{\infty}$ with the Spearman Rank 
correlation coefficient of 0.63, which can be well described by 
\begin{equation}
\log(\frac{\tau}{\rm days}) = 4.46\pm0.02 + (1.57\pm0.04)\log(\frac{SF_{\infty}}{\rm mag}) 
\end{equation}, 
based on considerations of the uncertainties in both coordinates. The 
determined power law function has the slope of $\sim1.57$ similar as 
the one in \citet{mi10}. Therefore, our determined parameters of 
$\ln(\tau)$ and $\ln(\sigma)$ should be reliable. Then, after 
considerations of the cosmological time dilation, the rest-frame $\tau$ 
and the rest-frame $\sigma$ are compared between the 106 
double-peaked emitters and the 7804 normal quasars and shown in 
Fig.~\ref{drw}. The much different distributions of the rest-frame 
$\ln(\tau)$ can be found and shown in the bottom panel of Fig.~\ref{drw}, 
with the mean value of the rest-frame $\ln(\tau/{\rm days})$ about 4.71  
for the normal quasars. And moreover, through the Student's T-statistic 
technique, the calculated T-statistic value and its significance are 15 and 
$2.8~\times~10^{-50}$ for the distributions of $\ln(\tau)$, respectively, 
which indicate significantly different mean values of $\ln(\tau)$ 
between the double-peaked emitters and the normal 
quasars with confidence level higher than 99.9\%.

\begin{figure*}
\centering\includegraphics[width = 18cm,height=10cm]{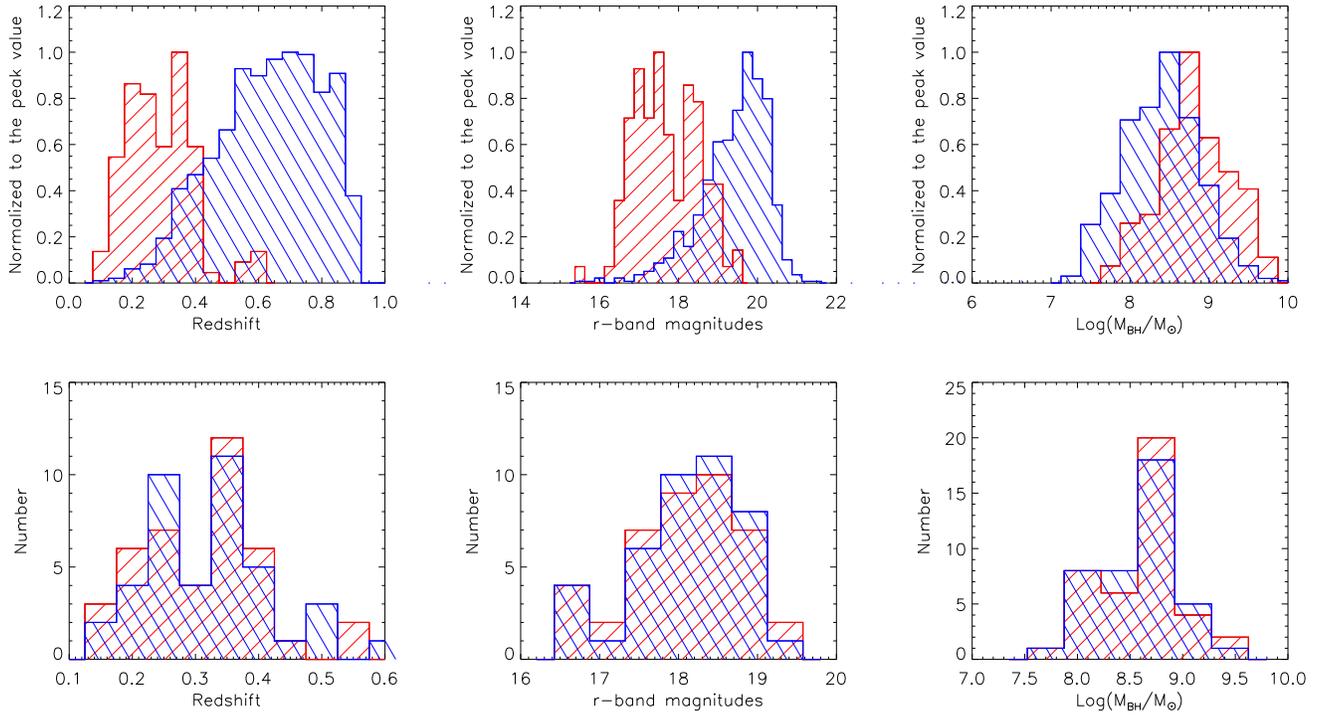}
\caption{Distributions of redshift (left panels), SDSS 
r-band magnitude (middle panels) and virial BH mass (right panels) of 
the double-peaked emitters and the normal quasars. In each 
panel, solid line in red and in blue represent the results for the 
double-peaked emitters and for the normal quasars, 
respectively. Top panels show the results with much different 
distributions for all the normal quasars and all the selected 
double-peaked emitters. Bottom panels show the results 
with the same distributions for the two subsamples including the 41 
double-peaked emitters and the 41 normal quasars.
}
\label{dis}
\end{figure*}

   In order to confirm the longer $\ln(\tau)$ in the 
double-peaked emitters, the following effects are mainly considered. 
First and foremost, we mainly consider the effects from different 
distributions of redshift, magnitude and BH mass between the 
double-peaked emitters and the normal quasars. As what we have shown 
in the introduction, the variability timescales should sensitively 
depend on BH masses. Therefore, if the selected double-peaked 
emitters had more massive black holes, the longer variability time 
scales could be expected. Moreover, some dependence of BH masses can 
be roughly found on the magnitudes and redshifts. Top panels of Fig.~\ref{dis}
show the much different distributions of redshift, magnitude and BH mass
between the double-peaked emitters and the normal quasars.  
Here, the virial BH masses of the double-peaked emitters 
and the normal quasars are collected from the \citet{sh11} based on 
the equation in \citet{vp06} (with the applied R-L relation similar to 
the more recent results in \citet{ben13}). And, in order to ignore 
effects of being applied different broad line widths to estimate 
BH masses, the virial BH masses estimated through broad H$\beta$ are  
only considered. There are 961 normal quasars and 79 double-peaked emitters 
with reliable virial BH masses collected from \citet{sh11}. For the 
other 27 double-peaked emitters not included in the \citet{sh11}, the 
virial BH masses are estimated through the parameters of broad H$\beta$ 
by the same equation in \citet{vp06} through the SDSS spectra, similar 
as what have been done in \citet{sh11}.  

\begin{figure*}
\centering\includegraphics[width = 16cm,height=10cm]{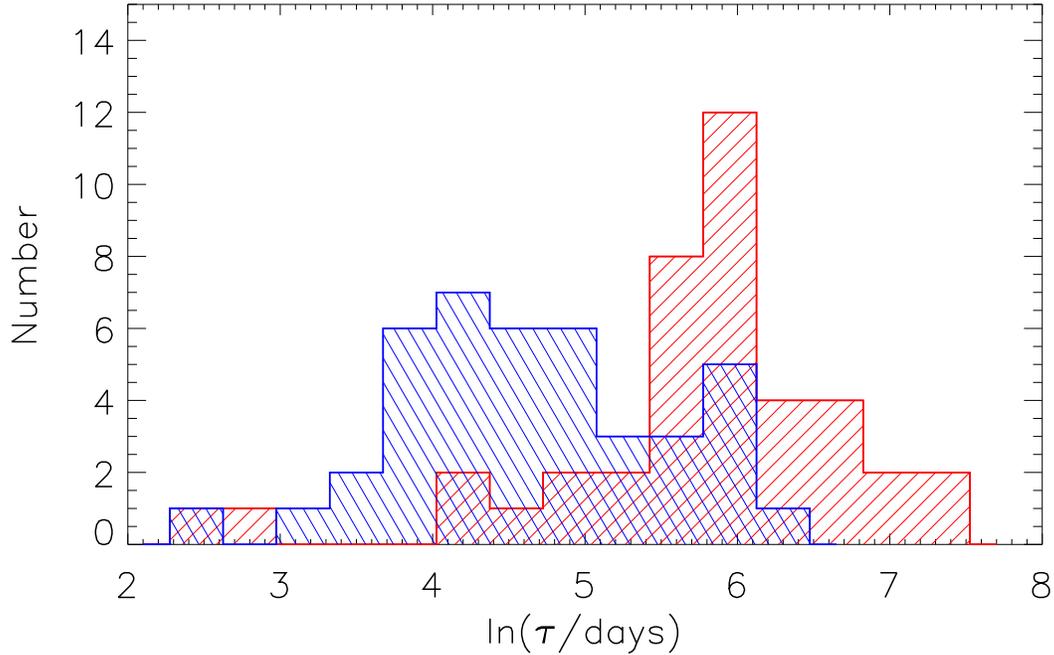}
\caption{Distributions of the rest-frame $\ln(\tau)$ for 
the 41 double-peaked emitters and for the 41 normal 
quasars shown in the bottom panels of Fig.~\ref{dis}. Solid line in red 
and solid line in blue represent the results for the 
double-peaked emitters and for the normal quasars, respectively.}
\label{dis2}
\end{figure*}

  
    In order to ignore the effects of the different distributions of 
redshift, magnitude and BH mass on the results in Fig.~\ref{drw} (longer 
$\ln(\tau)$ in the double-peaked emitters) as much as possible, the 
most convenient way is to create two subsamples for the double-peaked 
emitters and for the normal quasars, with the objects in the two 
subsamples having the same distributions of redshift, magnitude and BH 
mass. Based on the parameters of redshift, SDSS r-band magnitude and 
BH mass of the double-peaked emitters and the normal quasars, two 
subsamples including 41 double-peaked emitters and 41 normal quasars 
are created with the same distributions of the redshift, the SDSS r-band 
(with more higher signal-to-noises) apparent PSF magnitude and the virial 
BH mass, as the results shown in the bottom panels of Fig.~\ref{dis}. 
The two-sided Kolmogorov-Smirnov statistic technique is applied to check 
whether are there the same parameter distributions between the two samples. 
And we can find that the two distributions in each bottom panel of 
Fig.~\ref{dis} have the same distributions with confidence level higher 
than 90\%. Basic parameters including redshift, magnitude and BH mass are 
listed in Table~2 for the 41 double-peaked emitters and the 41 normal 
quasars in the subsamples. Then, properties of the DRW process determined 
$\ln(\tau)$ are compared between the 41 double-peaked emitters and the 
41 normal quasars. As shown in Fig.~\ref{dis2}, the mean values of the 
rest-frame $\ln(\tau/{\rm days})$ are about 5.72 and 4.63 days for 
the 41 double-peaked emitters and for the 41 normal quasars, respectively.

\begin{figure*}
\centering\includegraphics[width = 16cm,height=10cm]{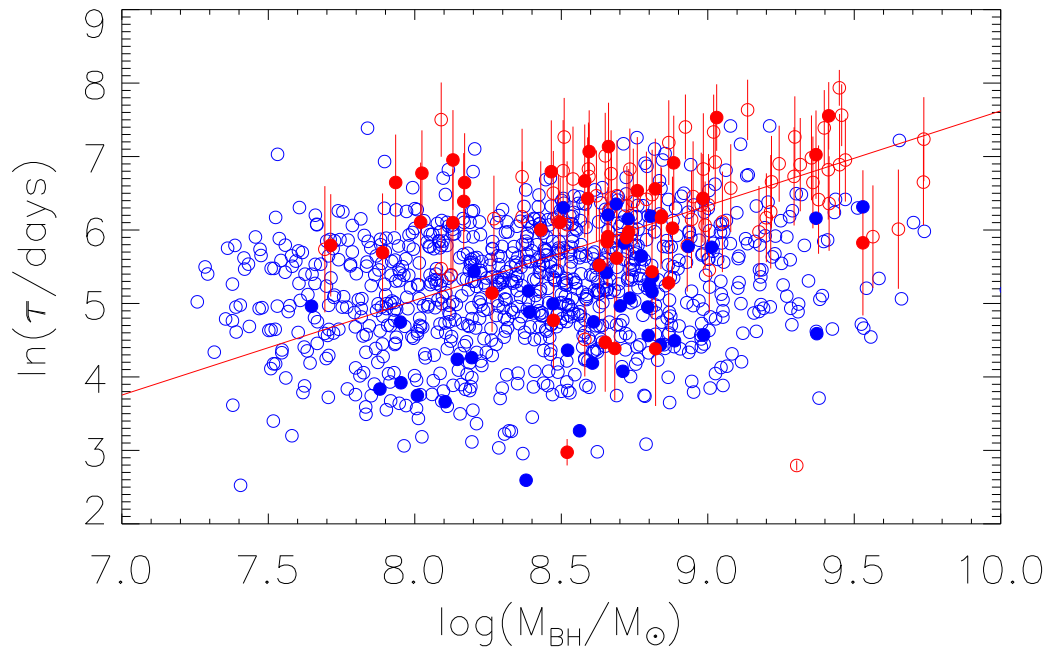}
\caption{Correlation between the observed-frame $\ln(\tau)$
and $M_{\rm BH}$. Circles in blue and circles in red with error bars are 
for the 961 normal quasars with reliable virial BH masses and for all the 
106 double-peaked emitters with reliable virial BH masses, respectively. 
Solid Circles in blue and solid circles in red with error bars are for 
the 41 normal quasars and for the 41 double-peaked emitters in the 
subsamples, respectively. The solid line in red shows the formula
$\tau\propto M_{\rm BH}^{0.56}$ reported in \citet{kbs09}.
}
\label{drw_mass}
\end{figure*}

  Moreover, as the results reported in \citet{kbs09}, 
Fig.~\ref{drw_mass} shows the dependence of $\ln(\tau)$ on BH mass for 
the double-peaked emitters and for the normal quasars in the parent 
samples and in the subsamples. And we can find that the reported 
formula $\tau\propto M_{\rm BH}^{0.56}$ in \citet{kbs09} can be well 
applied to describe the dependence shown in Fig.~\ref{drw_mass} based 
on our measured parameters. The results not only indicate that our 
measured parameters are reliable, but also illustrate that the 41 
normal quasars and the 41 double-peaked emitters in the subsamples 
have similar BH mass ranges as the objects in the parent samples, 
leading to the not much different distributions of $\ln(\tau)$ shown 
in the bottom panel of Fig.~\ref{drw} and in Fig.~\ref{dis2}. Then, 
the Student's T-statistic technique is applied again to the 
distributions of the rest-frame $\ln(\tau)$ of the 41 
double-peaked emitters and the 41 normal quasars. The calculated 
T-statistic value and its significance are 5.6 and $3.3~\times~10^{-7}$, 
respectively, which indicate significantly different mean values 
of $\ln(\tau)$ between the 41 double-peaked emitters 
and the 41 normal quasars with confidence level higher than 99.9\%. 
Thus, after well considerations of different redshift distributions, 
different magnitude distributions and different BH mass distributions, 
the more longer $\ln(\tau)$ can be re-confirmed in the 
double-peaked emitters.

   Besides, we should note that the SDSS g-filter for the normal 
quasars in SDSS S82 and the filter in the CSS for the 
double-peaked emitters covered different wavelength ranges, which 
would lead to some probable effects on the results shown in 
Fig.~\ref{drw} and in Fig.~\ref{dis2}. And moreover, the CSS DR2 
provided light curves and the SDSS S82 provided light curves have 
much different sampling rates and much different signal-to-noise ratios. 
Then, the probable effects are considered by two different methods as 
follows. The first method is to check magnitude correlation 
between the CSS DR2 provided magnitudes and the SDSS S82 provided 
magnitudes, to consider probable effects of different filters. The 
second method is to directly check the DRW process determined parameter 
correlations based on the CSS DR2 provided light curves and based on 
the SDSS S82 provided light curves for a sample of quasars, to 
consider probable effects of different sampling rates and different 
signal-to-noise ratios of light curves from the CSS DR2 and from the 
SDSS S82.

\begin{figure*}
\centering\includegraphics[width = 16cm,height=10cm]{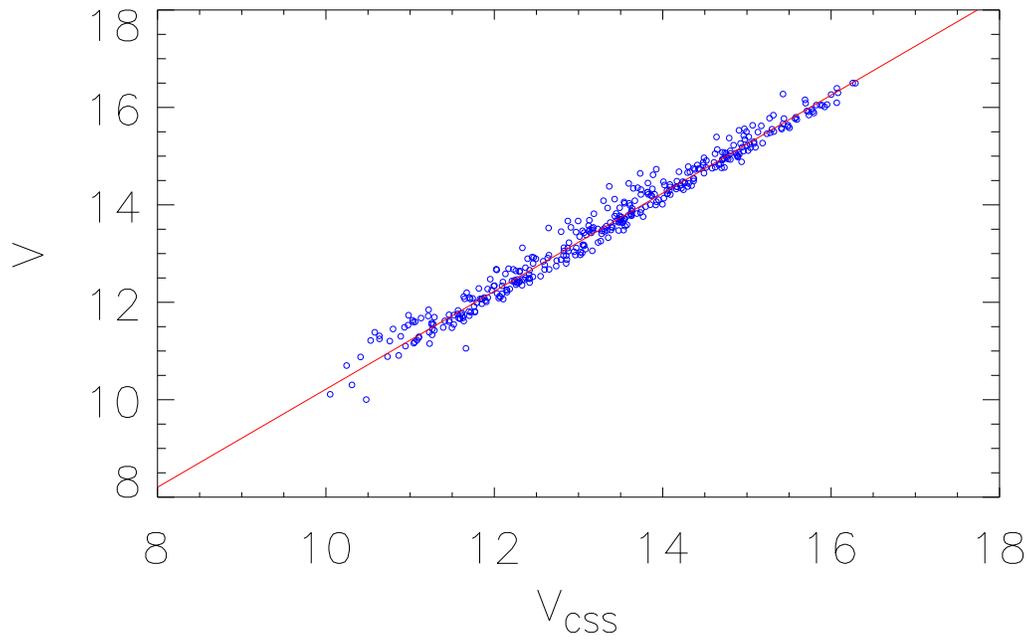}
\caption{Correlation between $V_{\rm CSS}$ 
and $V$ for more than 400 standard stars covered in the CSS.
The solid line in red shows the best fitted result by 
$V~=~0.1558~+~1.006 V_{\rm CSS}$.
}
\label{star}
\end{figure*}

   For the first method, properties of standard stars are considered as 
follows. Based on the standard stars from the catalogue of \citet{lan09} 
covered in the CSS, the correlation between the standard V band magnitudes 
$V$ and the CSS DR2 provided magnitudes $V_{\rm CSS}$ can 
be checked. Here, we do not list the magnitudes of $V$ and $V_{\rm CSS}$ 
of the more than 400 standard stars, but Fig~\ref{star} shows the 
correlation. The strong linear correlation with the Spearman rank 
correlation coefficient of 0.98 can be confirmed and described by 
\begin{equation} 
V~=~0.1558~+~1.006 V_{\rm CSS}
\end{equation}.
Moreover, \citet{je05} have shown that the standard V band magnitudes of 
SDSS quasars can be transferred from the SDSS g- and r-band magnitudes 
($g$ and $r$),
\begin{equation}
V~=~g~-~0.52(g~-~r)~-~0.03
\end{equation}.
Therefore, the SDSS g-band variability and CSS DR2 provided variability 
should have similar properties as the standard V band variability, 
to some extent.


   For the second method, among the 7804 normal quasars in the SDSS 
S82, there are 3207 normal quasars are also covered by the CSS. Then, the 
JAVELIN code is applied to determine the DRW process parameters of the 
CSS DR2 provided light curves of the 3207 normal quasars. Based on the 
DRW process determined parameters of $\ln(\tau)$ and $\ln(\sigma)$ at 
least 5 times larger than the corresponding uncertainties, there are 154 
normal quasars with well determined DRW process parameters from both the 
SDSS S82 provided light curves and from the CSS DR2 provided light curves. 
In top panels of Fig.~\ref{s82_css}, we show an example with light curves 
provided by both SDSS S82 and the CSS DR2, and their corresponding best 
fitted results by the JAVELIN code. Here, due to large magnitude 
uncertainties (small signal-to-noise ratios), the DRW process 
parameters can not be well determined 
from the CSS DR2 provided light curves for a large part of the 3207 
objects. In bottom panels of Fig.~\ref{s82_css}, we show an example with 
the CSS DR2 provided light curve which can not be described by the DRW 
process. Then, Fig.~\ref{drw3} shows the correlations of the DRW process 
parameters determined from the SDSS S82 provided light curves 
($\ln(\tau_{\rm SDSS})$ and $\ln(\sigma_{\rm SDSS})$) and from the CSS 
DR2 provided light curves ($\ln(\tau_{\rm CSS})$ and 
$\ln(\sigma_{\rm CSS})$) of the 154 quasars. The Spearman rank correlation 
coefficients are about 0.3 with $P_{null}<10^{-3}$ for the correlations 
of $\ln(\tau_{\rm SDSS})$ versus $\ln(\tau_{\rm CSS})$ and 
$\ln(\sigma_{\rm SDSS})$ versus $\ln(\sigma_{\rm CSS})$. The large 
scatters of the correlations shown in Fig.~\ref{drw3} are probably due 
to the different qualities of the light curves provided by the SDSS S82 
and by the CSS DR2. However, we can find that the ratio of 
$\tau_{\rm SDSS}$ to $\tau_{\rm CSS}$ has the mean value about 1.03, 
which can not be applied to explain the results shown in Fig.~\ref{drw} 
and in Fig.~\ref{dis2} that the mean $\tau$ ratio of the double-peaked 
emitters to the normal quasars is about 3. Therefore, even with 
considerations of the different qualities of the light curves applied 
to the normal quasars and to the double-peaked emitters, the longer 
$\ln(\tau)$ in the double-peaked emitters can be re-confirmed.


   Last but not the least, there is a point we should note that accretion 
rate plays an important role in AGN variability. If the double-peaked 
emitters had statistically higher accretion rates than the normal quasars, 
the longer variability timescales could be expected in the double-peaked 
emitters. Therefore, it is necessary to check properties of the accretion 
rates. Here, the dimensionless accretion rate ($\dot{M}$) is applied 
to trace the physical accretion rates ($\dot{m}$), 
\begin{equation} 
\dot{M}~=~\frac{L_{\rm bol}}{L_{\rm Edd}}~=~\frac{9\times~\lambda L_{\rm 5100\AA}/{\rm erg/s}}{1.38~\times~10^{38}~M_{\rm BH}/{\rm M_\odot}}
\end{equation},
where $L_{\rm Edd}$ and $L_{\rm bol}$ represent the Eddington luminosity and 
the bolometric luminosity simply estimated by the 9 times of the continuum 
luminosity at 5100\AA\ which can be collected from \citet{sh11} and/or from 
the SDSS spectra, respectively. The dimensionless accretion rates 
of the objects in the subsamples are listed in Table~2. The results are 
shown in Fig.~\ref{dotm}. And we can find that there are a bit smaller 
dimensionless accretion rates in the double-peaked emitters than in the 
normal quasars. The mean values of $\log(\dot{M})$ are about -1.52 and 
-1.64 for the 41 double-peaked emitters in the subsample and for all the 
106 double-peaked emitters, respectively. However, the mean values of 
$\log(\dot{M})$ are about -1.31 and -1.02 for the 41 normal quasars in 
the subsample and for the 961 normal quasars, respectively. In other words, 
the longer variability timescales in the double-peaked emitters are not 
due to larger accretion rates in the double-peaked emitters. 

   In brief, statistically longer DRW process determined intrinsic 
variability timescales can be found and confirmed in the 
double-peaked emitters, even after considerations of necessary effects, 
especially the effects from different distributions of redshift, 
BH mass and accretion rate. Therefore, rather than the observed 
double-peaked profiles of broad emission lines, probable intrinsic 
different structures of accretion disk could be expected in the 
double-peaked emitters. And hence, the 
double-peaked emitters can be well treated as an unique subclass of 
AGN with special properties of emission regions. 

   Before the end of the section, we try to find an interpretation of 
the longer variability timescales in the double-peaked 
emitters. By the expressions on the disk orbital or thermal timescales, 
BH mass and distance of emission regions to central BH are the two main 
parameters. As the results shown in Fig.~\ref{dis2} and in Fig.~\ref{dotm}, 
even the double-peaked emitters have the same BH masses 
and the similar accretion rates as the normal quasars, the longer 
variability timescales can be still confirmed. Therefore, the key point 
to explain the longer variability timescales is how to explain the 
longer distances of the emission regions to central BHs in the 
double-peaked emitters than in the normal quasars. 
Under the theoretical framework of standard thin accretion disk proposed 
by \citet{ss73}, the BH mass and the accretion rate can be effectively 
determine the radial dependence of temperature (the locations of emission 
regions). For the double-peaked emitters and the normal 
quasars as the results shown in Fig.~\ref{dis2} and in Fig.~\ref{dotm}, 
the same BH masses and similar accretion rates roughly indicating similar 
radial dependence of temperature (similar locations of emission regions) 
can not explain the longer variability timescales in the 
double-peaked emitters. Otherwise, radial dependence of accretion rate 
is considered, $\dot{m}_{\rm R}~\propto~R^{\beta}$, as supposed in 
\citet{lnr01} (see also the necessary modification of 
standard disk model for the well-known double-peaked emitter Arp 102B 
in \citet{po14}). Therefore, in the 
double-peaked emitters, simple radial dependence of accretion rate 
with larger $\beta$ could be an acceptable interpretation of larger 
distance of emission regions to central black hole, although it is not 
clear what mechanisms control the $\dot{m}_{\rm R}$ (radial flows?).

\section{Conclusions}

   Our main results and conclusions are as follows. First and foremost, 
the long-term light curves from the CSS DR2 have been well
analyzed by the DRW process for a large sample of SDSS spectroscopically
confirmed double-peaked emitters. Meanwhile, the long-term
light curves from the SDSS S82 have been analyzed by the same DRW process
for the spectroscopically confirmed normal quasars. Besides, the longer
variability timescales can be confirmed in the double-peaked
emitters than in the normal quasars, even after considerations of 
necessary effects, such as effects from different distributions of redshift, 
BH mass and accretion rate between the normal quasars and the 
double-peaked emitters. The radial dependence of accretion 
rate $\dot{m}_{\rm R}~\propto~R^{\beta}$ with larger value of $\beta$ 
could be an acceptable interpretation of the longer variability timescales 
in the double-peaked emitters. Last but not the least, the intrinsic 
difference between the double-peaked emitters and the 
normal quasars strongly supports that the double-peaked 
emitters could be well treated as a unique subclass of AGN.

\section*{Acknowledgements}
Zhang and FLL gratefully acknowledge the anonymous referee for 
giving us constructive comments and suggestions to greatly improve our paper.
Zhang acknowledges the kind support from the Chinese grant NSFC-U1431229.
FLL is supported under the NSFC grants 11273060, 91230115 and 11333008,
and State Key Development Program for Basic Research of China
(No. 2013CB834900 and 2015CB857000). This manuscript has made use of the
data from the CSS. The CSS survey is funded by the National Aeronautics
and Space Administration and the U.S. National Science Foundation. The
CSS website is http://nesssi.cacr.caltech.edu/DataRelease/.
This manuscript has made use of the data from the SDSS projects. Funding
for SDSS-III has been provided by the Alfred P. Sloan Foundation, the
Participating Institutions, the National Science Foundation, and the
U.S. Department of Energy Office of Science. The SDSS-III web site is
http://www.sdss3.org/. SDSS-III is managed by the Astrophysical Research
Consortium for the Participating Institutions of the SDSS-III
Collaboration including the University of Arizona, the Brazilian
Participation Group, Brookhaven National Laboratory, Carnegie Mellon
University, University of Florida, the French Participation Group, the
German Participation Group, Harvard University, the Instituto de
Astrofisica de Canarias, the Michigan State/Notre Dame/JINA
Participation Group, Johns Hopkins University, Lawrence Berkeley National
Laboratory, Max Planck Institute for Astrophysics, Max Planck Institute
for Extraterrestrial Physics, New Mexico State University, New York
University, Ohio State University, Pennsylvania State University,
University of Portsmouth, Princeton University, the Spanish Participation
Group, University of Tokyo, University of Utah, Vanderbilt University,
University of Virginia, University of Washington, and Yale University.

\begin{table*}
\centering
\caption{Basic Parameters of the 106 double-peaked emitters}
\begin{tabular}{cccc|cccc|cccc}
\hline\hline
name & z & $\ln(\tau)$ & $\ln(\sigma)$  & 
name & z & $\ln(\tau)$ & $\ln(\sigma)$  & 
name & z & $\ln(\tau)$ & $\ln(\sigma)$ \\
\hline
000710+005329  &  0.32  &  7.0$\pm$0.5  &  -1.6$\pm$0.2  &
001224$-$102226  &  0.22  &  5.9$\pm$0.7  &  -2.6$\pm$0.3  &
002444+003221  &  0.40  &  6.6$\pm$0.6  &  -2.2$\pm$0.2  \\
012128+003450  &  0.33  &  2.6$\pm$0.1  &  -1.4$\pm$0.1  &
013023+000551  &  0.34  &  5.1$\pm$0.6  &  -1.3$\pm$0.2  &
013253$-$095239  &  0.26  &  6.1$\pm$0.6  &  -1.7$\pm$0.2  \\
014901$-$080838  &  0.21  &  6.4$\pm$0.6  &  -2.1$\pm$0.2  &
015530$-$085704  &  0.16  &  6.3$\pm$0.6  &  -2.0$\pm$0.2  &
021259$-$003028  &  0.39  &  6.0$\pm$0.6  &  -2.1$\pm$0.2  \\
022930$-$000845  &  0.61  &  5.3$\pm$0.9  &  -2.0$\pm$0.3  &
030021$-$071459  &  0.39  &  6.6$\pm$0.5  &  -1.8$\pm$0.2  &
031235$-$063431  &  0.35  &  4.1$\pm$0.6  &  -2.1$\pm$0.2  \\
032559+000800  &  0.36  &  5.8$\pm$0.6  &  -1.9$\pm$0.2  &
034247+010933  &  0.36  &  5.7$\pm$0.5  &  -1.8$\pm$0.2  &
073927+404347  &  0.21  &  5.9$\pm$0.8  &  -2.4$\pm$0.4  \\
074157+275519  &  0.33  &  4.1$\pm$0.7  &  -2.0$\pm$0.1  &
074500+292702  &  0.33  &  5.8$\pm$0.6  &  -1.7$\pm$0.2  &
075403+481428  &  0.27  &  6.4$\pm$0.5  &  -1.8$\pm$0.2  \\
075408+431610  &  0.35  &  7.2$\pm$0.4  &  -1.5$\pm$0.2  &
080310+293234  &  0.33  &  7.3$\pm$0.4  &  -1.2$\pm$0.2  &
081051+174034  &  0.34  &  6.5$\pm$0.6  &  -1.9$\pm$0.2  \\
081916+481746  &  0.22  &  6.6$\pm$0.7  &  -2.1$\pm$0.3  &
082125+421907  &  0.22  &  5.3$\pm$0.7  &  -2.0$\pm$0.2  &
082406+334245  &  0.32  &  6.3$\pm$0.6  &  -1.8$\pm$0.2  \\
083225+370736  &  0.09  &  6.8$\pm$0.5  &  -1.2$\pm$0.2  &
083343+074654  &  0.21  &  6.7$\pm$0.5  &  -2.1$\pm$0.2  &
083413+590300  &  0.31  &  6.5$\pm$0.6  &  -1.6$\pm$0.3  \\
084110+022952  &  0.33  &  4.9$\pm$0.7  &  -2.1$\pm$0.2  &
084143+281956  &  0.47  &  5.5$\pm$0.6  &  -2.0$\pm$0.2  &
084205+075926  &  0.13  &  6.6$\pm$0.5  &  -2.2$\pm$0.2  \\
085029+185350  &  0.57  &  5.9$\pm$0.6  &  -2.1$\pm$0.2  &
085633+595747  &  0.28  &  6.4$\pm$0.6  &  -2.2$\pm$0.2  &
091706+325625  &  0.36  &  6.0$\pm$0.7  &  -2.7$\pm$0.3  \\
091828+513932  &  0.18  &  6.5$\pm$0.7  &  -2.7$\pm$0.3  &
091930+110854  &  0.37  &  5.6$\pm$0.6  &  -2.4$\pm$0.2  &
091941+534549  &  0.56  &  6.0$\pm$0.7  &  -2.2$\pm$0.3  \\
092133+321320  &  0.58  &  6.8$\pm$0.5  &  -2.1$\pm$0.2  &
092659+093248  &  0.42  &  5.8$\pm$0.6  &  -2.2$\pm$0.2  &
093509+481909  &  0.22  &  5.5$\pm$0.6  &  -2.2$\pm$0.2  \\
093653+533127  &  0.23  &  6.6$\pm$0.6  &  -2.4$\pm$0.3  &
093756+104809  &  0.27  &  2.5$\pm$0.0  &  -0.7$\pm$0.0  &
093844+005715  &  0.17  &  7.2$\pm$0.4  &  -1.9$\pm$0.2  \\
094215+090016  &  0.21  &  5.5$\pm$0.8  &  -2.7$\pm$0.2  &
095236+205143  &  0.28  &  4.5$\pm$0.6  &  -2.3$\pm$0.1  &
095939+480439  &  0.39  &  6.0$\pm$0.8  &  -2.4$\pm$0.3  \\
100027+025951  &  0.34  &  7.0$\pm$0.5  &  -1.5$\pm$0.2  &
101931+262642  &  0.25  &  6.2$\pm$0.6  &  -1.5$\pm$0.2  &
103059+310256  &  0.18  &  5.4$\pm$0.6  &  -2.0$\pm$0.2  \\
103202+600836  &  0.29  &  5.7$\pm$0.9  &  -1.6$\pm$0.2  &
103427+614821  &  0.16  &  6.5$\pm$0.7  &  -2.2$\pm$0.2  &
103620+121734  &  0.19  &  6.8$\pm$0.5  &  -1.7$\pm$0.2  \\
105041+345631  &  0.27  &  5.5$\pm$0.6  &  -2.3$\pm$0.2  &
105115+280527  &  0.42  &  5.8$\pm$0.6  &  -2.2$\pm$0.2  &
105624+601558  &  0.20  &  5.9$\pm$0.7  &  -1.6$\pm$0.2  \\
110051+170934  &  0.35  &  6.0$\pm$0.6  &  -2.0$\pm$0.2  &
110920+213802  &  0.34  &  6.8$\pm$0.5  &  -1.6$\pm$0.2  &
111230+181311  &  0.19  &  6.4$\pm$0.6  &  -2.3$\pm$0.3  \\
111537+542725  &  0.42  &  5.6$\pm$0.8  &  -2.0$\pm$0.2  &
111916+110107  &  0.39  &  4.0$\pm$0.7  &  -1.7$\pm$0.1  &
112007+423551  &  0.23  &  4.3$\pm$0.5  &  -1.3$\pm$0.1  \\
113021+022211  &  0.24  &  5.6$\pm$0.5  &  -2.1$\pm$0.2  &
113640+573840  &  0.15  &  5.5$\pm$0.8  &  -2.4$\pm$0.3  &
114335$-$002942  &  0.17  &  5.6$\pm$0.6  &  -2.5$\pm$0.2  \\
115227+604818  &  0.27  &  6.4$\pm$0.7  &  -1.5$\pm$0.2  &
115408+252145  &  0.34  &  4.8$\pm$0.5  &  -2.1$\pm$0.2  &
120442+275412  &  0.16  &  6.0$\pm$0.5  &  -2.1$\pm$0.2  \\
120714+292236  &  0.38  &  6.2$\pm$0.5  &  -1.6$\pm$0.2  &
120924+103611  &  0.39  &  6.4$\pm$0.6  &  -2.7$\pm$0.3  &
121037+315705  &  0.39  &  7.2$\pm$0.4  &  -1.9$\pm$0.2  \\
121716+080942  &  0.34  &  6.6$\pm$0.6  &  -2.1$\pm$0.2  &
123054+110010  &  0.23  &  6.8$\pm$0.5  &  -2.1$\pm$0.2  &
123215+132033  &  0.28  &  5.6$\pm$0.6  &  -2.0$\pm$0.2  \\
123807+532556  &  0.35  &  6.9$\pm$0.5  &  -1.3$\pm$0.2  &
123945+195425  &  0.24  &  6.4$\pm$0.6  &  -2.1$\pm$0.2  &
125142+240435  &  0.19  &  6.1$\pm$0.5  &  -1.7$\pm$0.2  \\
125809+351943  &  0.29  &  7.6$\pm$0.2  &  -0.8$\pm$0.1  &
130634+304934  &  0.41  &  5.8$\pm$0.7  &  -2.1$\pm$0.2  &
130927+032251  &  0.26  &  6.3$\pm$0.6  &  -2.0$\pm$0.3  \\
132145+033056  &  0.27  &  6.1$\pm$0.5  &  -1.5$\pm$0.2  &
132834$-$012917  &  0.15  &  6.6$\pm$0.5  &  -2.0$\pm$0.2  &
133053+311930  &  0.24  &  5.8$\pm$0.6  &  -1.8$\pm$0.2  \\
133329+154550  &  0.25  &  5.2$\pm$0.5  &  -2.1$\pm$0.2  &
133433$-$013825  &  0.29  &  5.7$\pm$0.8  &  -2.5$\pm$0.3  &
134548+114443  &  0.13  &  5.2$\pm$0.5  &  -2.3$\pm$0.2  \\
135529+352332  &  0.31  &  6.9$\pm$0.5  &  -1.7$\pm$0.2  &
135712+170444  &  0.15  &  6.5$\pm$0.6  &  -2.1$\pm$0.2  &
140336+174136  &  0.23  &  6.0$\pm$0.7  &  -2.0$\pm$0.2  \\
141613+021908  &  0.16  &  7.1$\pm$0.5  &  -2.6$\pm$0.2  &
141628+124213  &  0.33  &  6.1$\pm$0.7  &  -2.0$\pm$0.3  &
142522+080326  &  0.23  &  6.2$\pm$0.6  &  -1.9$\pm$0.2  \\
142725+194952  &  0.11  &  6.2$\pm$0.7  &  -1.7$\pm$0.2  &
143204+394437  &  0.35  &  6.1$\pm$0.7  &  -2.5$\pm$0.2  &
151132+100953  &  0.28  &  6.0$\pm$0.5  &  -1.7$\pm$0.2  \\
152139+033729  &  0.13  &  6.3$\pm$0.5  &  -1.8$\pm$0.2  &
154019$-$020505  &  0.32  &  6.6$\pm$0.5  &  -2.0$\pm$0.2  &
154433+202626  &  0.27  &  6.7$\pm$0.6  &  -2.4$\pm$0.3  \\
161218+073145  &  0.21  &  5.9$\pm$0.6  &  -2.2$\pm$0.2  &
161811+093052  &  0.26  &  5.7$\pm$0.7  &  -2.1$\pm$0.2  &
163856+433512  &  0.34  &  7.2$\pm$0.4  &  -1.2$\pm$0.2  \\
165822+183735  &  0.17  &  6.6$\pm$0.6  &  -2.5$\pm$0.2  &
170102+340400  &  0.09  &  5.3$\pm$0.5  &  -2.4$\pm$0.2  &
172711+632242  &  0.21  &  5.8$\pm$0.8  &  -2.0$\pm$0.3  \\
212501$-$081328  &  0.62  &  6.4$\pm$0.6  &  -2.3$\pm$0.2  &
212619$-$065408  &  0.42  &  7.1$\pm$0.5  &  -1.9$\pm$0.2  &
222024+010931  &  0.21  &  6.6$\pm$0.5  &  -1.8$\pm$0.2  \\
233254+151305  &  0.21  &  6.2$\pm$0.7  &  -2.8$\pm$0.3  & 
               &         &               &                &
               &         &               &                \\  
\hline\hline
\end{tabular}\\
Notice: Col(1), Col(5) and Col(9) list the object name in the format of 
SDSS Jhhmmss$\pm$ddmmss, Col(2), Col(6) and Col(10) list the information 
of redshift, Col(3), Col(7) and Col(11) list the determined DRW parameter 
of rest-frame $\ln(\tau)$ in unit of days, Col(4), Col(8) and Col(12) list the 
determined DRW parameter of rest-frame $\ln(\sigma)$ in unit of 
${\rm mag/day^{0.5}}$. 
\end{table*}

\begin{table*}
\centering
\caption{Basic Parameters of the 41 double-peaked emitters  
and the 41 normal quasars in the subsamples}
\begin{tabular}{cccccc|cccccc}
\hline\hline
name & z & mag & $\log(M_{\rm BH})$ & $\dot{M}$ & $\ln(\tau)$ & 
name & z & mag & $\log(M_{\rm BH})$ & $\dot{M}$ & $\ln(\tau)$ \\
\hline
\multicolumn{6}{c|}{Double-peaked emitters}  &  
\multicolumn{6}{c}{Normal quasars} \\
\hline
012128+003450  & 0.25  & 19.08  & 8.51  & -2.01  & 2.6  &  000557+002837  & 0.33  & 18.41  & 8.20  & -1.30  & 5.2  \\ 
013023+000551  & 0.37  & 18.70  & 8.80  & -1.61  & 5.1  &  001030+010006  & 0.33  & 17.49  & 8.93  & -1.27  & 5.4  \\ 
014901$-$080838  & 0.25  & 18.58  & 8.17  & -2.04  & 6.4  &  002831$-$000413  & 0.26  & 17.66  & 8.71  & -1.48  & 3.8  \\ 
031235$-$063431  & 0.25  & 18.65  & 8.65  & -1.41  & 4.1  &  003723+000812  & 0.34  & 18.34  & 7.95  & -0.97  & 4.5  \\ 
032559+000800  & 0.34  & 17.74  & 9.05  & -2.28  & 5.8  &  004458+004319  & 0.37  & 18.67  & 8.65  & -1.42  & 5.9  \\ 
073927+404347  & 0.32  & 18.51  & 8.02  & -2.01  & 5.9  &  010151+002028  & 0.23  & 18.52  & 8.80  & -1.84  & 4.9  \\ 
074157+275519  & 0.34  & 18.38  & 8.68  & -1.44  & 4.1  &  010230$-$003206  & 0.33  & 18.99  & 8.83  & -1.68  & 4.1  \\ 
074500+292702  & 0.34  & 17.99  & 8.49  & -1.23  & 5.8  &  011402$-$004750  & 0.33  & 18.68  & 8.19  & -1.00  & 3.9  \\ 
082125+421907  & 0.34  & 19.02  & 8.63  & -2.51  & 5.3  &  012050$-$001832  & 0.22  & 18.61  & 9.37  & -2.07  & 4.2  \\ 
083413+590300  & 0.40  & 18.17  & 8.46  & -1.25  & 6.5  &  012602$-$000822  & 0.31  & 17.96  & 8.60  & -1.05  & 3.8  \\ 
084110+022952  & 0.33  & 19.05  & 8.86  & -2.02  & 4.9  &  013015+002557  & 0.34  & 19.05  & 8.52  & -1.37  & 4.0  \\ 
084143+281956  & 0.30  & 19.43  & 8.65  & -1.21  & 5.5  &  013352+011345  & 0.49  & 18.21  & 8.38  & -1.11  & 4.9  \\ 
084205+075926  & 0.39  & 17.34  & 8.02  & -1.48  & 6.6  &  013418+001536  & 0.14  & 16.79  & 8.80  & -0.85  & 5.8  \\ 
085029+185350  & 0.44  & 17.80  & 8.98  & -1.16  & 5.9  &  013923$-$010632  & 0.52  & 18.86  & 8.88  & -1.52  & 4.1  \\ 
091941+534549  & 0.33  & 18.89  & 8.75  & -1.20  & 6.0  &  014017$-$005003  & 0.58  & 16.43  & 9.01  & -1.02  & 5.4  \\ 
092659+093248  & 0.33  & 17.88  & 8.84  & -1.10  & 5.8  &  015105$-$003426  & 0.40  & 18.87  & 8.80  & -1.71  & 5.2  \\ 
093756+104809  & 0.24  & 17.86  & 9.30  & -1.99  & 2.5  &  024052$-$004110  & 0.24  & 17.84  & 9.16  & -2.03  & 4.3  \\ 
095236+205143  & 0.20  & 17.59  & 8.47  & -1.11  & 4.5  &  024601$-$005937  & 0.27  & 18.13  & 8.10  & -1.22  & 3.4  \\ 
095939+480439  & 0.28  & 17.97  & 8.60  & -1.91  & 6.0  &  031142$-$005918  & 0.40  & 19.11  & 8.37  & -1.38  & 2.3  \\ 
103059+310256  & 0.26  & 16.76  & 8.68  & -1.43  & 5.4  &  205212$-$002645  & 0.21  & 18.63  & 8.57  & -1.49  & 4.3  \\ 
103202+600836  & 0.48  & 19.22  & 8.43  & -2.28  & 5.7  &  205608+002928  & 0.28  & 17.45  & 9.16  & -1.13  & 5.9  \\ 
105041+345631  & 0.58  & 18.62  & 7.71  & -0.68  & 5.5  &  210839$-$004816  & 0.28  & 19.08  & 8.81  & -1.33  & 4.9  \\ 
105115+280527  & 0.28  & 17.24  & 9.20  & -1.98  & 5.8  &  211838+004035  & 0.48  & 18.60  & 7.64  & -0.62  & 4.7  \\ 
105624+601558  & 0.14  & 18.23  & 8.13  & -1.58  & 5.9  &  213110$-$003537  & 0.20  & 17.61  & 8.00  & -1.23  & 3.6  \\ 
110051+170934  & 0.23  & 18.60  & 8.16  & -0.95  & 6.0  &  213245+000146  & 0.34  & 18.43  & 7.95  & -0.99  & 3.7  \\ 
110920+213802  & 0.34  & 18.20  & 8.66  & -1.24  & 6.8  &  213818+011222  & 0.36  & 17.55  & 8.39  & -0.85  & 4.5  \\ 
111230+181311  & 0.33  & 18.12  & 7.93  & -1.02  & 6.4  &  215010$-$001000  & 0.25  & 18.38  & 8.77  & -1.53  & 5.3  \\ 
111537+542725  & 0.21  & 18.25  & 8.73  & -1.10  & 5.6  &  215408$-$002744  & 0.42  & 17.88  & 8.71  & -1.64  & 5.6  \\ 
113021+022211  & 0.27  & 17.63  & 8.72  & -1.50  & 5.6  &  215949+001004  & 0.25  & 17.98  & 8.47  & -1.28  & 4.7  \\ 
113640+573840  & 0.36  & 17.32  & 7.89  & -1.65  & 5.5  &  220332+004401  & 0.15  & 18.17  & 8.72  & -1.18  & 5.8  \\ 
114335$-$002942  & 0.40  & 17.67  & 8.65  & -1.76  & 5.6  &  221155$-$001043  & 0.22  & 18.13  & 8.79  & -1.20  & 4.6  \\ 
115227+604818  & 0.49  & 19.02  & 8.58  & -2.29  & 6.4  &  221737$-$004854  & 0.26  & 19.52  & 8.65  & -1.35  & 5.0  \\ 
115408+252145  & 0.21  & 17.58  & 8.26  & -0.64  & 4.8  &  222024+010931  & 0.34  & 16.61  & 8.73  & -1.18  & 4.8  \\ 
121037+315705  & 0.22  & 16.65  & 9.03  & -0.95  & 7.2  &  222230+010231  & 0.33  & 18.70  & 8.56  & -1.72  & 3.0  \\ 
121716+080942  & 0.26  & 16.76  & 8.88  & -1.11  & 6.6  &  230007+001739  & 0.39  & 18.54  & 8.14  & -1.13  & 4.0  \\ 
123054+110010  & 0.23  & 16.62  & 8.59  & -1.05  & 6.8  &  230155$-$010649  & 0.24  & 16.82  & 8.68  & -1.23  & 6.1  \\ 
130634+304934  & 0.42  & 18.64  & 8.84  & -1.33  & 5.8  &  230705+004754  & 0.44  & 18.63  & 8.70  & -1.37  & 4.6  \\ 
130927+032251  & 0.15  & 18.73  & 8.82  & -1.98  & 6.3  &  232259$-$005359  & 0.32  & 17.18  & 7.88  & -0.85  & 3.6  \\ 
154433+202626  & 0.33  & 18.20  & 8.13  & -1.09  & 6.7  &  232525+000352  & 0.25  & 18.10  & 8.50  & -1.32  & 6.0  \\ 
161811+093052  & 0.52  & 17.73  & 8.87  & -1.71  & 5.7  &  234145$-$004640  & 0.27  & 17.62  & 8.98  & -1.17  & 4.1  \\ 
163856+433512  & 0.27  & 18.34  & 9.41  & -1.98  & 7.2  &  235457+004219  & 0.34  & 17.87  & 8.79  & -1.48  & 4.3  \\ 
\hline\hline
\end{tabular}\\
Notice: The first six columns show the parameters of the object name in 
the format of SDSS Jhhmmss$\pm$ddmmss, redshift, SDSS r-band magnitude, 
virial BH mass in unit of ${\rm M_\odot}$, dimensionless accretion rate 
$\dot{M}$ and the rest-frame $\ln(\tau)$ in unit of days for the 41 
double-peaked emitters in the subsample. The last six columns shows 
the corresponding parameters for the 41 normal quasars in the subsample.
\end{table*}

\clearpage

\begin{figure*}
\centering\includegraphics[width = 18cm,height=12cm]{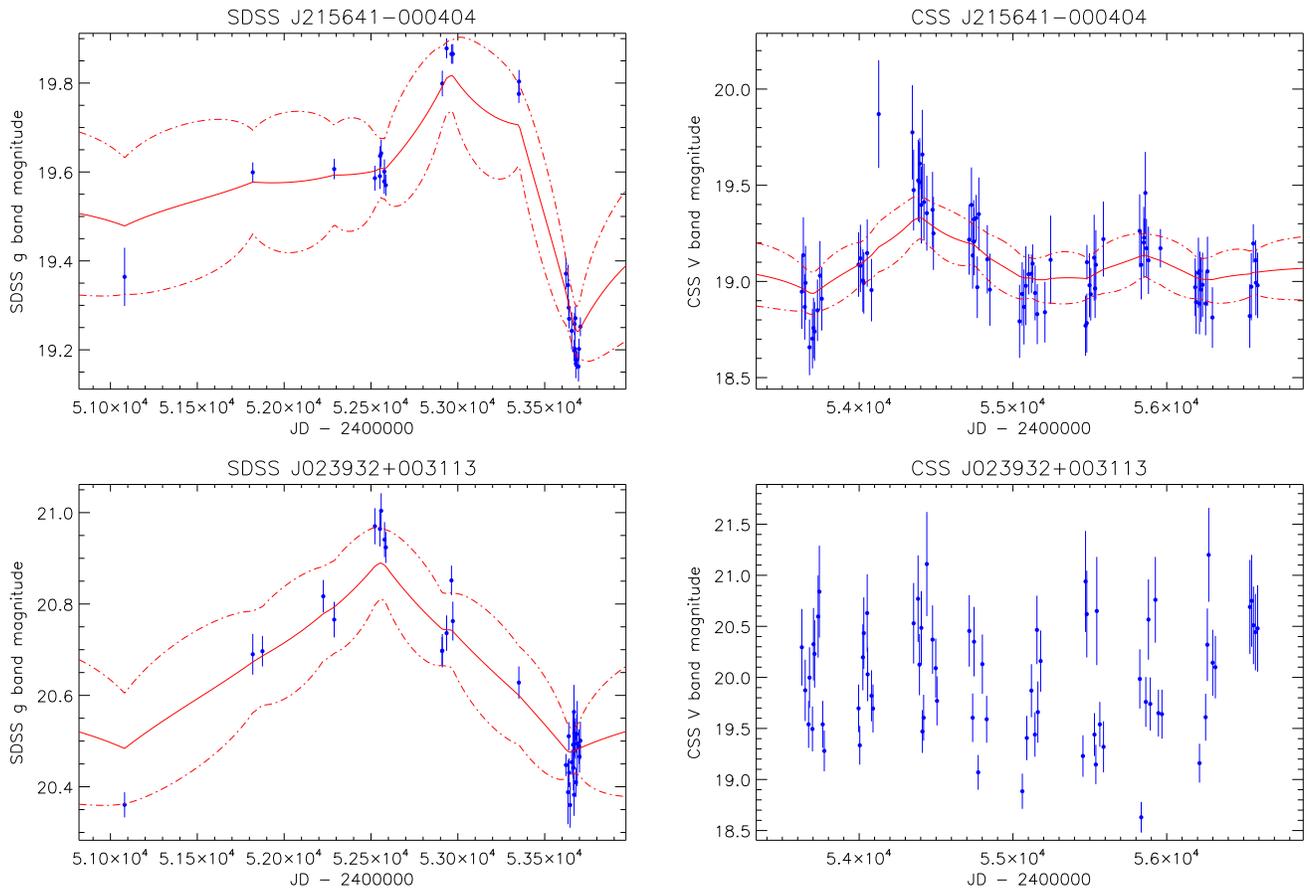}
\caption{Properties of the SDSS S82 provided g-band light 
curves (in the left panels) and the CSS DR2 provided light curves (in the 
right panels) of two normal quasars. In the panels, solid circles in blue 
are for the observed light curves, solid lines in red and dot-dashed lines 
in red show the best fitted results by the DRW process and the 
corresponding 1sigma confidence bands. In the bottom right panel, due to 
large magnitude uncertainties, the light curve can not be well described 
by the DRW process.
}
\label{s82_css}
\end{figure*}

\begin{figure*}
\centering\includegraphics[width = 16cm,height=20cm]{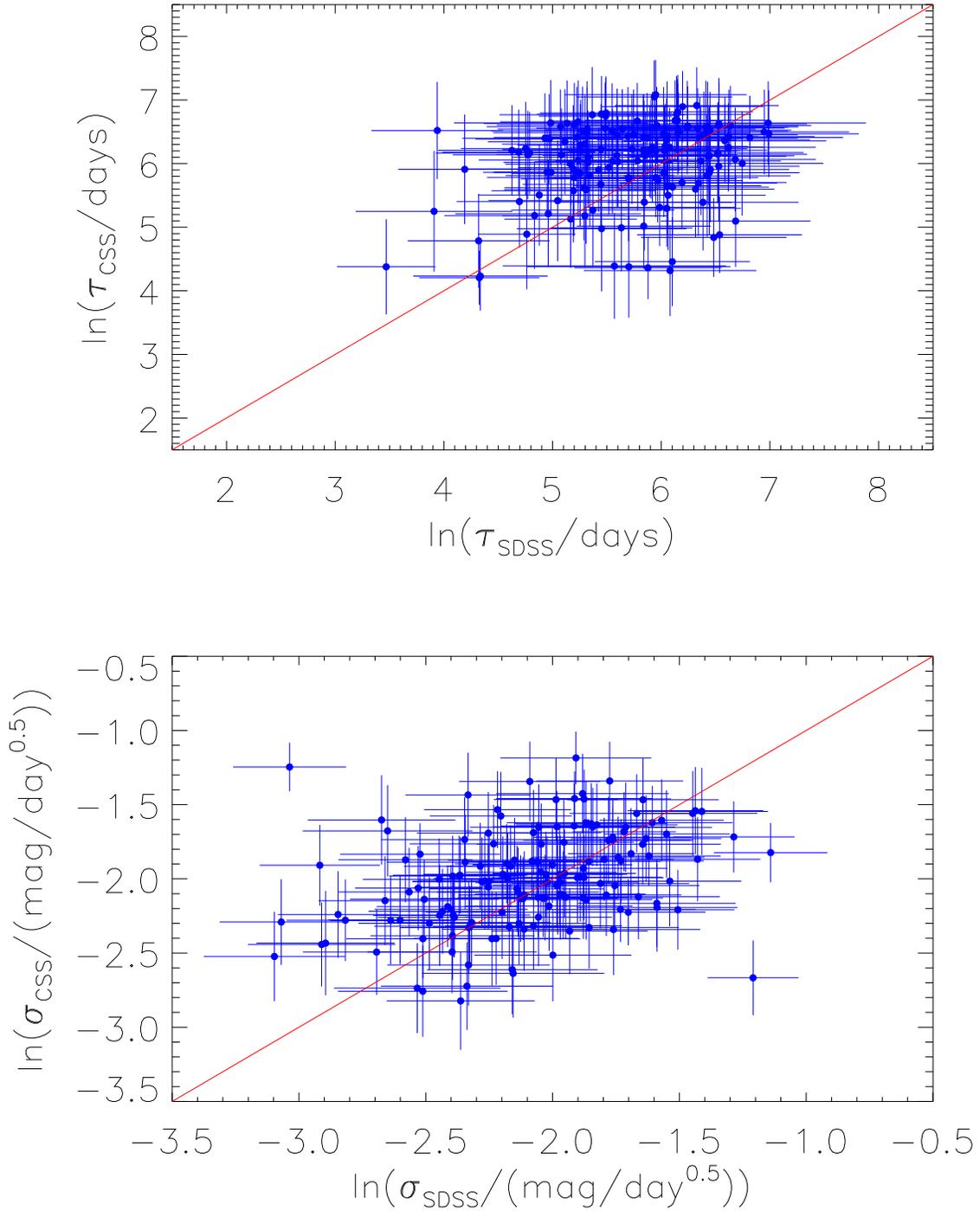}
\caption{Correlation between the observed-frame 
$\ln(\tau_{\rm SDSS})$ and the observed-frame $\ln(\tau_{\rm CSS})$ (top panel) 
and the correlation between the observed-frame $\ln(\sigma_{\rm SDSS})$ and the
observed-frame $\ln(\sigma_{\rm CSS})$ (bottom panel). In each panel, the 
solid line in red shows X=Y.
}
\label{drw3}
\end{figure*}

\begin{figure*}
\centering\includegraphics[width = 16cm,height=20cm]{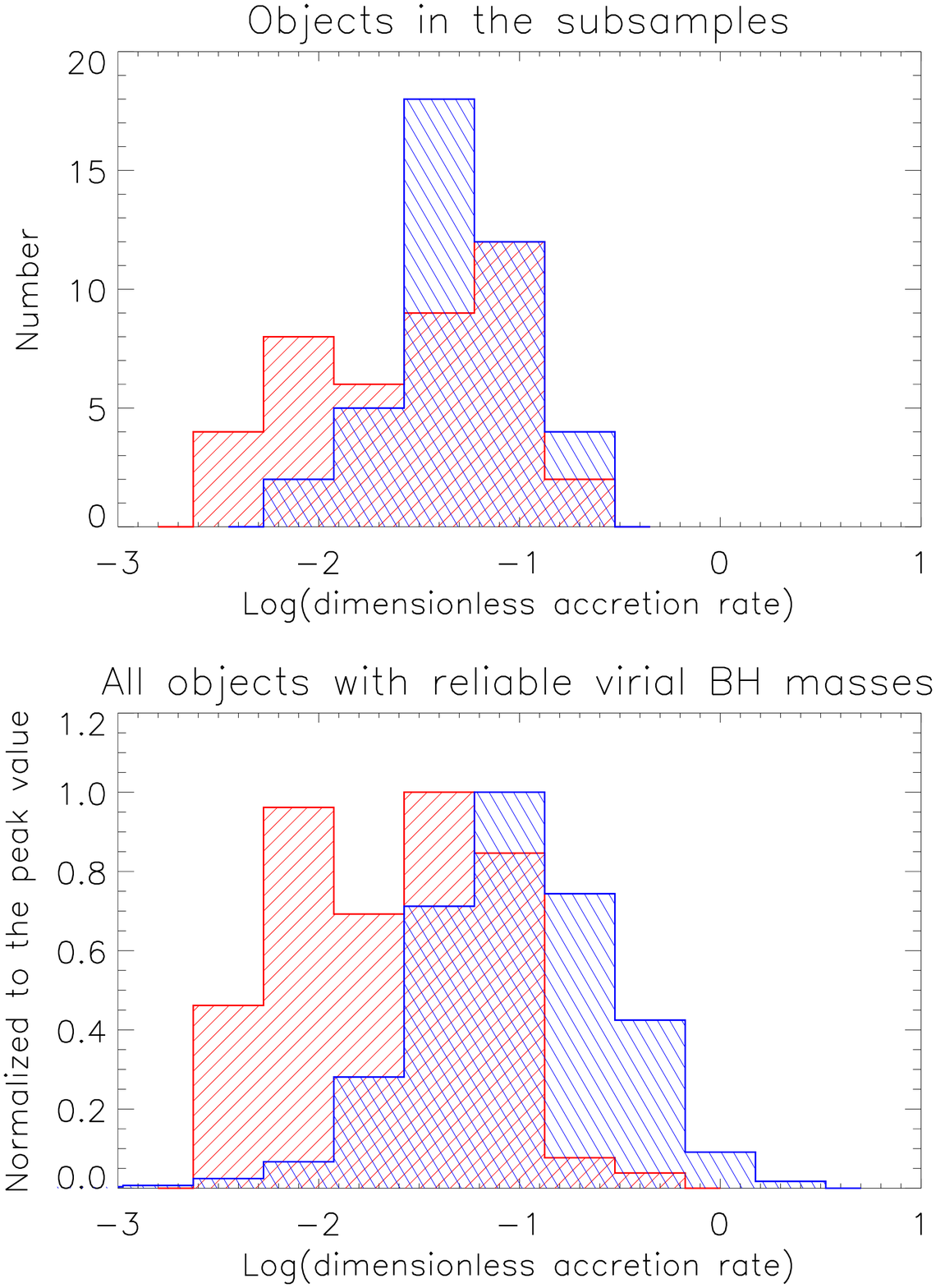}
\caption{Distributions of the dimensionless
accretion rates. Top panel shows the results for the 41 
double-peaked emitters (in red) and for the 41 normal 
quasars (in blue) in the subsamples. Bottom panel shows the results 
for all the 106 double-peaked emitters (in red) and for the 961 normal 
quasars (in blue) with reliable virial BH masses.
}
\label{dotm}
\end{figure*}

\label{lastpage}
\end{document}